\documentclass[aps,prl,twocolumn,showpacs,superscriptaddress,final]{revtex4-2}

\usepackage{amsmath}
\usepackage{amssymb}
\usepackage{graphicx}
  \usepackage{hyperref}
\usepackage[arrow, matrix, curve]{xy}
\usepackage{bm}
\usepackage{yhmath}
\usepackage{color}
\usepackage{url}
\newcommand\diff{\mathrm{d}}

\renewcommand\vec[1]{\boldsymbol{\mathrm{#1}}}

\usepackage[usenames,dvipsnames]{xcolor}
\hypersetup{colorlinks=true, linkcolor=BrickRed, urlcolor=blue!50!black, citecolor=blue!50!black}

\usepackage[normalem]{ulem}

\makeatletter
\newcommand\hide@visible[1]{%
  \bgroup\fboxsep=.3ex\colorbox{Gray}{begin hide}%
  #1\colorbox{Gray}{end hide}\egroup%
}
\newcommand\hide@hidden[1]{%
  \bgroup\fboxsep=.3ex\colorbox{Gray}{hidden text}%
}
\newcommand\hide@invisible[1]{}
\newcommand\makevisible{\let\hide\hide@visible}
\newcommand\makehidden{\let\hide\hide@hidden}
\newcommand\makeinvisible{\let\hide\hide@invisible}
\newcommand{\Var}{\mathrm{Var}}
\makeatother
\makehidden

\begin{document}

\texttt{Published: Physical Review E 112, L013401 (2025)}
\title{Glass transition in colloidal monolayers controlled by light-induced caging
}

\author{Abolfazl Ahmadirahmat}
\affiliation{Institut f{\"u}r Theoretische Physik,  Universit\"at Innsbruck, Technikerstra{\ss}e 25/2, A-6020 Innsbruck, Austria}

\author{Michele Caraglio}
\affiliation{Institut f{\"u}r Theoretische Physik,  Universit\"at Innsbruck,
Technikerstra{\ss}e 25/2, A-6020 Innsbruck, Austria}
\author{Vincent Krakoviack}
\affiliation{ENS de Lyon, CNRS, Laboratoire de Chimie (LCH UMR5182) et Centre Blaise Pascal, 69342 Lyon cedex 07, France}

\author{Thomas Franosch}
\affiliation{Institut f{\"u}r Theoretische Physik, Universit\"at Innsbruck,
Technikerstra{\ss}e 25/2, A-6020 Innsbruck, Austria}

\email[]{thomas.franosch@uibk.ac.at}

\date{\today}

\begin{abstract}

We theoretically 
 
investigate the glass-transition problem for a  
quasi-two-dimensional colloidal dense suspension  modulated by a one-dimensional periodic external potential as imposed by interfering laser beams. 
 Relying on a mode-coupling approach, we examine the nonequilibrium state diagram for hard disks  as a function of the density and the period of the modulation for various potential strengths.  

The competition between the local packing and the distortion of the cages induced by the potential leads to a striking nonmonotonic behavior of the glass-transition line which allows  melting of a glass state merely by changing the external fields. 
In particular, we find regions in the non-equilibrium state diagram where a moderate periodic modulation stabilizes the liquid state.  

\end{abstract}


\maketitle

Understanding how the dynamics of atoms, molecules, or colloidal
particles is influenced by an external heterogeneous potential energy
landscape, either deterministic or random, has been a major
goal of classical statistical physics~\cite{Haus:PhysRep_150:1987,
  Havlin:AdvPhys_36:1987, Bouchaud:PhysRep_195:1990}.  
Indeed, besides being a fundamental question, this problem is highly relevant for several practical situations, such as confined transport in ordered or amorphous porous media~\cite{Karger:DiffNanoMat:2012, Huber:JPhysCondMatt_27:2015, Sun:JPhysChemLett_11:2020},  and diffusion over corrugated or rough surfaces~\cite{Choi:IECR_40:2001,Antczak:SurfDiff:2010}.
Thus, it encompasses different research domains, including chemical and material engineering, biophysics and geophysics.

In many important cases, periodicity, i.e., discrete translational
invariance, is present in some or all directions of space.  For example,
porous solids such as zeolites~\cite{Baerlocher:zeoliteatlas:2007, Kellouai:JChemPhys_160:2024} are
well described as 3D-periodic interconnected networks of pores or,
more abstractly, as 3D-periodic arrays of obstacles.  An atom  adsorbed at the
surface of a crystal is exposed to a 2D-periodic potential energy
landscape due to the arrangement of the atoms in the subjacent solid~\cite{Barth:SurfSciRep_40:2000}.  In the colloidal domain,
1D-periodicity is achieved in monolayers, with either paramagnetic
particles over a magnetically structured stripe-patterned substrate~\cite{Tierno:PCCP_42:2009} or polarizable particles plunged in the
light field generated by interfering laser beams~\cite{Dalle-Ferrier:SM_7:2011,
  Jenkins:JoP_40:2008,Capellmann:JCP_148:2018, Castaneda:SM_21:2025}.  In the former case, a
potential for technological applications has been demonstrated, as the
setup allows particle separation and sorting and controlled transport
of micron-sized cargos~\cite{Tierno:PCCP_42:2009}.  Although periodic
light fields have not been explicitly considered in this respect,
analogous proofs of concept exist based on random-interference speckle
patterns~\cite{Volpe:SciRep_4:2014, *Volpe:OptExp_22:2014}.

The present theoretical study is motivated by the latter
2D systems with a 1D external periodic
potential modulation, as they represent minimal models with coexisting
continuous and discrete translational invariance.  Specifically, we
focus on  colloidal monolayers in a periodic light field, in which
the period of the potential can be made comparable to the diameter of
the colloids while the amplitude can be tuned between zero and several
$k_\mathrm{B} T$ by changing the laser intensities.  Hence, it is
possible to investigate how their equilibrium dynamics progressively
departs from that of the unperturbed state, see e.g. Refs.~\cite{Dalle-Ferrier:SM_7:2011, Stoop:PRL_124:2020,  Lips:CommunPhys_4:2021}.

Manipulation of colloidal particles with  periodic light fields has a long history, dating back to the pioneering works of Ashkin and collaborators~\cite{Ashkin:PRL_24:1970, Ashkin:Science_210:1980, Ashkin:PNAS_94:1997, Smith:OptLett_6:1981}, and later by Chowdhury \textit{et al.}~\cite{Chowdhury:PRL_55:1985}.  Close to the two-dimensional
bulk crystallization, a strong enough potential with a suitable period
can drive a colloidal fluid through a laser-induced freezing
transition, resulting in a near-hexagonal solid
phase~\cite{Chowdhury:PRL_55:1985, LOUDIYI19921,*LOUDIYI199226}. Further increases in modulation strength can result in laser-induced melting of the crystalline phase~\cite{Chakrabarti:PRL_75:1995, Wei:PRL_81:1998,Strepp:PRE_63:2001, Bechinger:PRL_86:2001, Kraft:PRE_102:2020}. 
Besides liquid-solid phase transformations in the system, the interplay between local packing and the pinning effect of the external modulation is naturally expected to have a nontrivial influence on the geometry of the cages, and consequently, on the dynamics of the liquid phase and the eventual liquid-glass transition, even at lower densities.

Yet, the \emph{dynamic behavior} of 
such modulated dense colloidal monolayers 
has not been explored so far, except for few simulation studies, e.g. Ref. ~\cite{Herrera-Velarde:PRE_79:2009, Pal:JCondMat_32:2019}. 
A systematic variation of the  two experimental 'knobs', period and amplitude of the modulation, promises to result in a deeper insight into the role of the cages~\cite{Li:Nature_587:2020, Zhang:PRL_132:2024, Barbhuiya:PRE_110:2024}. This strategy is similar in spirit to adding polymers to the suspension resulting in a new type of caging accompanied by a reentrant transition~\cite{Bergenholtz:PRE_59:1999, Fabbian:PRE_59:1999, Dawson:PRE_63:2000, Pham:Science_296:2002, Zaccarelli:JPhys_19:2007}, or confining a liquid in a slit to induce multiple reentrants~\cite{Lang:PRL_105:2010, *Lang:PRE_86:2012, Mandal:NatComm_5:2014,Mandal:SM_13:2017}, using mixtures of differently sized particles~\cite{Hajnal:PRE_80:2009, Voigtmann:EPL_96:2011}, or a porous environment~\cite{Krakoviack:PRL_94:2005,  *Krakoviack:PRE_75:2007, *Krakoviack:PRE_79:2009, Kurzidem:PRL_103:2009, Kim:EPL_88:2009}.

The goal of this Letter is  to investigate the ramifications of the modulation on the glassy dynamics of dense colloidal monolayers. 
 Our approach relies on a mode-coupling theory (MCT)~\cite{Goetze:Complex_Dynamics, Janssen:2018} 
suitably adapted for the broken translational symmetry along the modulation, see the   companion paper~\cite{Ahmadirahmat:PREL:2025} for details. 
 We then evaluate the theory numerically for hard disks to obtain 
the nonequlibrium  state diagram and discuss the emergence of non-monotonic
glass-transition lines as a result of the competition of local packing
and the modulation.

\textit{Theory.--} The colloidal monolayer is comprised of $N$ interacting particles enclosed in an area $A$ at number density $n_{0}=N/A$. 
The system is exposed to an external periodic potential $\mathcal{U}(z) = \mathcal{U}(z+a)$ where $z$ is the coordinate in the  direction of the modulation and $a$ its period. 
The external potential breaks the translational invariance along the $z$-direction, however a uniform shift of all particles by lattice vectors 
$\vec{R}\in \Lambda := \{ \vec{r} + n a \vec{e}_z : \vec{r}\perp \vec{e}_z, n \in \mathbb{Z}\}$ leaves the system invariant in a statistical sense. 
The associated reciprocal lattice $\Lambda^* = \{ \vec{Q}_\mu = (2\pi \mu/a) \vec{e}_z: \mu \in\mathbb{Z} \}$ is degenerate and consists of a one-dimensional lattice only. 
Any wave vector can be uniquely expressed as $\vec{q}+ \vec{Q}_\mu$  where $\vec{q} \in \text{BZ}:= \{\vec{q}: -\pi/a< \vec{q}\cdot \vec{e}_z \leq \pi/a\}$ is in the first Brillouin zone and $\vec{Q}_\mu \in \Lambda^*$ is a reciprocal lattice vector. 
We introduce fluctuating density modes 

\begin{align}\label{eq:densities}
\delta \rho_{\mu}(\vec{q},t) = \sum_{n=1}^N  e^{i ( \vec{q}+\vec{Q}_\mu )\cdot \vec{x}_n(t) } - A n_\mu \delta_{\vec{q},0},  
\end{align}
such that its canonical average $\langle \delta \rho_\mu(\vec{q},t) \rangle =0$ vanishes,  
where $n_\mu = \int_0^a n(z) \exp( i Q_\mu z) \diff z/a$  is the Fourier coefficient of the density modulation $n(z)$. From these density modes we construct generalized intermediate scattering functions (ISF)
\begin{align}
S_{\mu\nu}(\vec{q},t) = \frac{1}{N} \langle \delta \rho_\mu(\vec{q},t)^* \delta \rho_\nu(\vec{q},0) \rangle,
\end{align}
which encode the structural relaxation of the modulated liquid. The diagonal elements $\mu=\nu$ are just the conventional ISF corresponding to wave vector $\vec{q}+\vec{Q}_\mu$, while the off-diagonal elements allow for \emph{Umklapp} processes where the wave vectors of the density modulations differ by a reciprocal lattice vector $\vec{Q}_\mu-\vec{Q}_\nu\in \Lambda^*$.  In the companion paper~\cite{Ahmadirahmat:PREL:2025} we show that the generalized ISF provide information equivalent to the density-density correlation function in real space and derive further properties and symmetries. 

We use the Mori-Zwanzig projection operator formalism~\cite{Goetze:Complex_Dynamics, Hansen:Theory_of_Simple_Liquids} to reformulate the problem in terms of memory kernels. To keep the derivation simple, we rely here on Newtonian dynamics anticipating that the slow structural dynamics close to the glass transition is independent of the microscopic dynamics. The derivation is along the lines of the case of a confined liquid~\cite{Lang:PRL_105:2010, *Lang:PRE_86:2012} but is adapted to the symmetries of the underlying problem.
Choosing the densities of  Eq.~\eqref{eq:densities} as distinguished variables one arrives at the equation of motion
\begin{align}
\dot{S}_{\mu\nu}(\vec{q},t)  + &\sum_{\kappa\lambda\in \mathbb{Z} } \int_0^t K_{\mu\kappa}(\vec{q},t-t') \nonumber \\
& \times [\mathbf{S}^{-1}(\vec{q})]_{\kappa\lambda} S_{\lambda \nu}(\vec{q},t') \diff t' = 0 .
\end{align}
Here $\mathbf{S}(\vec{q})$ denotes the matrix of static structure factors with components $[\mathbf{S}(\vec{q})]_{\mu\nu} = S_{\mu\nu}(\vec{q},0)$ and the memory kernel $K_{\mu\nu}(\vec{q},t)$ is a correlation function  composed of particle currents.

By the particle conservation law
\begin{align}
\partial_t \delta \rho_\mu(\vec{q}, t) = i \sum_{\alpha=\parallel, \perp} (q^\alpha+Q_\mu^\alpha) j_\mu^\alpha(\vec{q},t),
\end{align}
the time derivative of the density is coupled to particle currents that split naturally into components 
\begin{subequations}
\begin{align}
j^\parallel_\mu(\vec{q},t) &= \sum_{n=1}^N \frac{ \hat{\vec{q}}^\parallel \cdot \vec{p}_n(t) }{m} \exp[ i (\vec{q}+\vec{Q}_\mu)\cdot \vec{x}_n(t) ] , \\
j^\perp_\mu(\vec{q},t) &= \sum_{n=1}^N \frac{ \hat{\vec{q}}^\perp \cdot \vec{p}_n(t) }{m} \exp[ i (\vec{q}+\vec{Q}_\mu)\cdot \vec{x}_n(t) ] ,
\end{align}
\end{subequations}
 parallel and perpendicular to the modulation. Here $m$ denotes the mass of a particle and $\hat{\vec{q}}^\parallel =\vec{e}_z$ is the direction of the modulation and $\hat{\vec{q}}^\perp = \vec{q}^\perp/ |\vec{q}^\perp|$ is the direction of the perpendicular component $\vec{q}^\perp = \vec{q}- \vec{e}_z  (\vec{q} \cdot \vec{e}_z)$ of the wave vector.   
The splitting of the currents suggests also decomposing the current kernel 
 \begin{align}
K_{\mu\nu}(\vec{q},t) = \sum_{\alpha\beta= \parallel,\perp} (q^\alpha+Q_\mu^\alpha) \mathcal{K}_{\mu\nu}^{\alpha\beta}(\vec{q},t) (q^\beta+ Q_\nu^\beta),
\end{align} 
where  the indices $\alpha, \beta$ are referred to as channel indices. 

Performing another projection using the currents $j^\alpha_\mu(\vec{q})$ as distinguished variables yields the exact equations of motion (for details see companion paper~\cite{Ahmadirahmat:PREL:2025})
\begin{align}
\partial_t \mathcal{K}^{\alpha\beta}_{\mu\nu}(\vec{q},t)    +  & \sum_{\gamma\delta=\parallel,\perp}\sum_{\kappa\lambda\in \mathbb{Z}}\int_0^t
{\mathcal{K}}^{\alpha\gamma}_{\mu\kappa}(\vec{q})
 \mathcal{M}^{\gamma\delta}_{\kappa\lambda}(\vec{q},t-t')  \nonumber \\ 
& \times   \mathcal{K}_{\lambda\nu}^{\delta\beta}(\vec{q},t') \diff t' =0 .
\end{align}

Here the static current matrix
\begin{align}
\mathcal{K}_{\mu\nu}^{\alpha\beta}(\vec{q}) := \mathcal{K}_{\mu\nu}^{\alpha\beta}(\vec{q},0)= \frac{1}{N} \langle j_\mu^\alpha(\vec{q})^* j_\nu(\vec{q}) \rangle = \frac{k_B T}{m n_{0}} \delta^{\alpha\beta} n_{\nu-\mu},  
\end{align}
has been introduced, while the nontrivial dynamics is hidden in the force kernel $\mathcal{M}^{\gamma\delta}_{\kappa\lambda}(\vec{q},t)$.

To obtain closed equations of motion, we rely on a mode-coupling approximation suitably generalized to the case of split currents. The goal is to approximate the memory kernel as bilinear functional of the matrix $\mathbf{S}(\vec{q},t)$  of the intermediate scattering function itself,  
$\mathcal{M}^{\alpha\beta}_{\mu\nu}(\vec{q},t)  
\approx\mathcal{F}_{\mu\nu}^{\alpha\beta}[\mathbf{S}(t),\mathbf{S}(t);\vec{q}]$. The mode-coupling functional $\mathcal{F}_{\mu\nu}^{\alpha\beta}$ is constructed by standard methods and relies on static input quantities only. After some calculations we find (for details see companion paper~\cite{Ahmadirahmat:PREL:2025})
\begin{align}\label{eq:effective}
\mathcal{M}^{\alpha\beta}_{\mu\nu}(\vec{q},t)  
&\approx\mathcal{F}_{\mu\nu}^{\alpha\beta}[\mathbf{S}(t),\mathbf{S}(t);\vec{q}] \nonumber\\
&=\frac{1}{2N} 
\sum_{\vec{q}_1\vec{q}_2 \in \text{BZ}} 
\sum_{\substack{\mu_1\mu_2 \\ \nu_1\nu_2}\in \mathbb{Z}} 
\mathcal{Y}^{\alpha}_{\mu;\mu_{1}\mu_{2}}(\vec{q},\vec{q}_{1}\vec{q}_{2})\nonumber\\
&\times S_{\mu_{1}\nu_{1}}(\vec{q}_{1},t) S_{\mu_{2}\nu_{2}}(\vec{q}_{2},t) \mathcal{Y}^{\beta}_{\nu;\nu_{1}\nu_{2}}(\vec{q},\vec{q}_{1}\vec{q}_{2})^*, 
\end{align}
where the coupling strength is encoded in the vertices
\begin{align}\label{eq:Y_vertices}
&\mathcal{Y}^{\alpha}_{\mu,\mu_{1}\mu_{2}}(\vec{q},\vec{q}_{1}\vec{q}_{2})\nonumber \\
=& n_{0}^2  \sum_{\mu^\sharp = 0, \pm 1} 
  \delta_{\vec{q}-\vec{Q}_{\mu^\sharp} ,\vec{q}_{1}+\vec{q}_{2}}  \sum_{\kappa\in \mathbb{Z} } v^*_{\mu-\kappa}
  \nonumber \\
  &\times  \left[ (q_1^\alpha+ Q_{\kappa+\mu^\sharp-\mu_2}^\alpha)
    c_{\kappa+ \mu^\sharp-\mu_2,\mu_1}(\vec{q}_1) 
+( 1 \leftrightarrow 2) \right]
.
\end{align}
Structural information enters in terms of the Fourier coefficients $v_\mu$ of the local volume $v(z) :=1/n(z)$ and the mode decomposition of the direct correlation function $c_{\mu\nu}(\vec{q})$. The latter are obtained by the mode decomposition of the Ornstein-Zernike relation   
\begin{align}
[\mathbf{S}^{-1}(\vec{q}) ]_{\mu\nu} = n_{0} v_{ \nu-\mu} - n_{0} c_{\mu\nu}(\vec{q})  .
\end{align}
The peculiar feature of the mode-coupling functional is that it allows for processes such that momentum is nonconserved. The sum of the wave vectors of the
ISF $\vec{q}_1,\vec{q}_2 \in \text{BZ}$ is not necessarily again in the first Brillouin zone but needs to be folded back by a reciprocal lattice vector. 
Correspondingly only the \emph{crystal momentum} is conserved as is familiar from solid state physics~\cite{Ashcroft:Solid_State_Physics:1976}.

To locate the glass transition point one has to solve the
self-consistent set of equations for the non-ergodicity parameters
$F_{\mu\nu}(\vec{q}) := \lim_{t\to\infty} S_{\mu\nu}(\vec{q},t) $:
\begin{align}\label{eq:Nonergodicity}
\mathbf{F}(\vec{q}) =& \mathbf{S}(\vec{q}) - [ \mathbf{S}^{-1}(\vec{q})+ \mathbf{G}^{-1}(\vec{q}) ]^{-1} \, , \\
G_{\mu\nu}(\vec{q}) =& \sum_{\alpha,\beta = \parallel, \perp} 
 (q^\alpha+Q_\mu^\alpha) [ \bm{\mathcal{N}}^{-1}(\vec{q})]^{\alpha\beta}_{\mu\nu}  (q^\beta+ Q_\nu^\beta), \nonumber
\end{align}
where $\bm{\mathcal{N}}(\vec{q}) $ is the long-time limit of the force kernel  $\bm{\mathcal{N}}(\vec{q}) = \lim_{t\to \infty} \bm{\mathcal{M}}(\vec{q},t) = \bm{\mathcal{F}}[\mathbf{F},\mathbf{F};\vec{q}] $. 
Glassy states are characterized by non-vanishing nonergodicity parameters $F_{\mu\nu}(\vec{q})  \neq 0$, while liquid states correspond to $F_{\mu\nu}(\vec{q}) = 0$. In a nonequilibrium-state diagram,  glassy and liquid regions emerge depending on  the control parameters, 
the glass-transition line then separates glassy from liquid states.

\textit{Results.--} We investigate the glass-transition line for hard disks of diameter $\sigma$, modulated by an external sinusoidal potential $\mathcal{U}(z) = U_1 \cos(2\pi z/a)$, where $U_{1}$ is the amplitude of the potential and $a$ is the period. The case without modulation has  been discussed in Ref.~\cite{Bayer:PRE_76:2007}. 
To numerically solve Eq.~\eqref{eq:Nonergodicity}, the total wave vectors $\vec{k} =\vec{q}+\vec{Q}_\mu$, $\vec{q} \in \text{BZ}, \vec{Q}_\mu \in \Lambda^*$.   have been discretized according to a uniform $2$D Cartesian grid, $\vec{k}\in \{ (n_x \Delta k, n_z \Delta k): n_x,n_z = 0,1,\ldots,N_k-1\}$, where $\Delta k = k_{\text{max}} / N_k$ is the discretization step, starting from a high-$k$ cut-off $k_{\text{max}}$ and the number of grid points in each direction $N_k$.
As a  compromise between  accuracy  and  numerical complexity, we have chosen $N_k = 270$ and  $k_{\text{max}} = 80.0/\sigma$ and carefully checked, for a  subset in parameter space, that for higher values of $k_{\text{max}}$ and $N_k$, the differences in the final results are negligible~\cite{Caraglio2020AnII}.

Additionally, to reduce numerical complexity, we have employed a diagonal approximation (DA) on the mode indices $\mu,\nu$, 
where the off-diagonal elements of $S_{\mu \nu} (\vec{q})$, $c_{\mu \nu}(\vec{q})$, $F_{\mu \nu}(\vec{q})$, and $\mathcal{F}_{\mu \nu}^{\alpha \beta} [\mathbf{F}, \mathbf{F};\vec{q}]$ for $\mu \neq \nu$ are set  to zero. Similar diagonal approximations 
have  already  been successfully adopted in other extensions of MCT~\cite{Franosch:PRE56:1997,Scheidsteger:PRE56:1997,Lang:PRL_105:2010,Mandal:NatComm_5:2014, *Mandal:SM_13:2017,Jung:PRE102:2020}.
The diagonal approximation can be overcome in principle, however, we anticipate  the numerical results not  to change the qualitative behavior, 
 while  only slight quantitative shifts are expected to  occur, as has been demonstrated recently for  slit geometry~\cite{Jung:PRE107:2023}					.
\begin{figure}
\includegraphics[width=\linewidth]{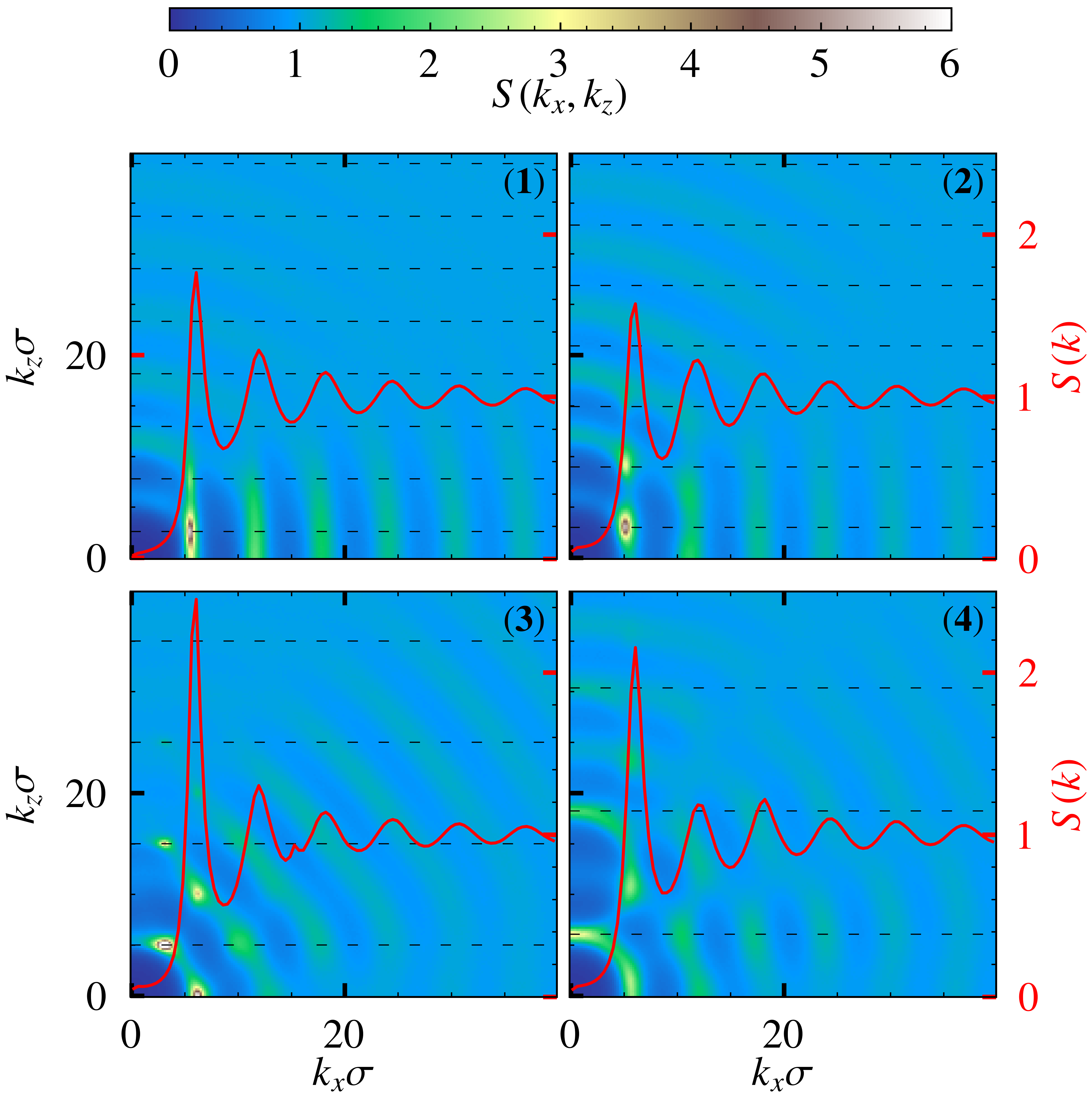}
\caption{\label{fig:ssf} Static structure factor $S(k_x, k_y)$ for hard disks at dimensionless density $n_{0} \sigma^2 =0.7$ 
and potential modulation amplitude $U_1 = 1.7k_{B}T$ for various periods: \textbf{(1)} $a=1.22\sigma$, \textbf{(2)} $a= 1.05\sigma$, \textbf{(3)} $a= 0.63\sigma$ and, \textbf{(4)} $a= 0.52\sigma$. Here, $\vec{k} = \vec{q} + \vec{Q}_\mu, \vec{q}\in BZ, \vec{Q}_\mu \in \Lambda^*$ is the total wave vector.
The full red lines are angular averages of the static structure factor, and the dashed black lines indicate the different  Brillouin zones.
}
\end{figure}

\begin{figure}[htp]
\includegraphics[width=\linewidth]{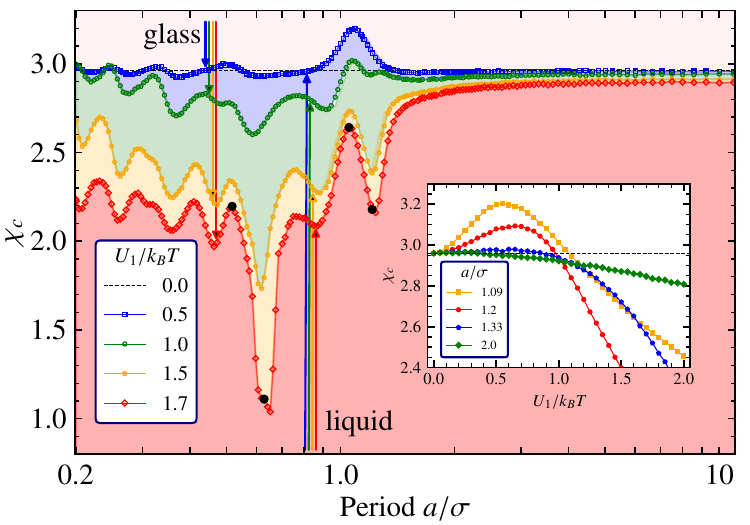}
\caption{\label{fig:phase} Nonequilibrium-state diagram:  The glass-transition line $\chi_c = \chi_c(a, U_1; \bar{\varphi})$ separating liquid from glassy states for various amplitudes $U_1$ within the diagonal 
approximation. Black dots on the red curve  correspond  to the reported static structure factors in Fig.~\ref{fig:ssf}. Inset:  Critical glass transition parameter $\chi_c(a, U_1; \bar{\varphi} )$ as a function of the amplitude $U_1$ for several  values of the period $a$  highlighting the suppression of  the glass transition. }
\end{figure}

The static structure factor, providing the input for the MCT equations, has been computed using Monte-Carlo simulations  for hard disks and different modulation amplitudes and periods. 
For a fixed amplitude $U_1$ of the external modulation, a glass transition occurs at some critical packing fraction $\varphi_c(a, U_1)$ as a function of the period $a$.
Collecting enough statistics from Monte-Carlo simulations to accurately obtain the static structure factor is computationally expensive.
Therefore, rather than varying the packing fraction $\varphi := n_{0} \pi \sigma^2 /4$,  we follow the strategy of Refs.~\cite{Sciortino:PRE_57:1998,Lang:PRL_105:2010} of a \emph{semi-schematic model} where a multiplicative control parameter $\chi$ for the MCT functional (evaluated at fixed reference packing fraction $\bar{\varphi}$) mimics the variation in density. 
The glass-transition line $\chi_c = \chi_c(a, U_1; \bar{\varphi})$ then separates liquid from glassy states as a function of the period $a$ in the nonequilbrium-state diagram for fixed amplitude $U_1$. 
We anticipate that the variation of $\varphi_c(a, U_1)$  is qualitatively well described by the $a$-dependence of the critical parameter $\chi_c(a, U_1; \bar{\varphi})$. 

In particular, in our Monte-Carlo simulation, we have chosen a reference hard-disk density of $n_{0} \sigma^2 = 0.7$ corresponding to a packing fraction $\bar{\varphi} \approx 0.55$.
For this density, by analyzing the static and dynamic properties from the Monte-Carlo simulations, we have checked that, for potential amplitudes $U_1 \lesssim 1.7 k_B T$, the system consistently maintains a liquid phase irrespective of the modulation period $a$.
This observation is in good agreement with previous studies on modulated colloidal liquids~\cite{Chakrabarti:PRL_75:1995, Wei:PRL_81:1998,Strepp:PRE_63:2001,Bechinger:PRL_86:2001,Kraft:PRE_102:2020}. 
Changing the period $a$ from $1.22\sigma$ to $0.52\sigma$  results in a non-monotonic dependence of the first sharp diffraction peak of the angular-averaged static structure factor, see Fig.~\ref{fig:ssf}.
The effect of the modulation becomes even more
pronounced, when inspecting the full wave-vector dependent structure
factor; in which case, particularly large peaks emerge for an external
potential of period $a = 0.63\sigma$.

\begin{figure}[htp!]
\includegraphics[width=\linewidth]{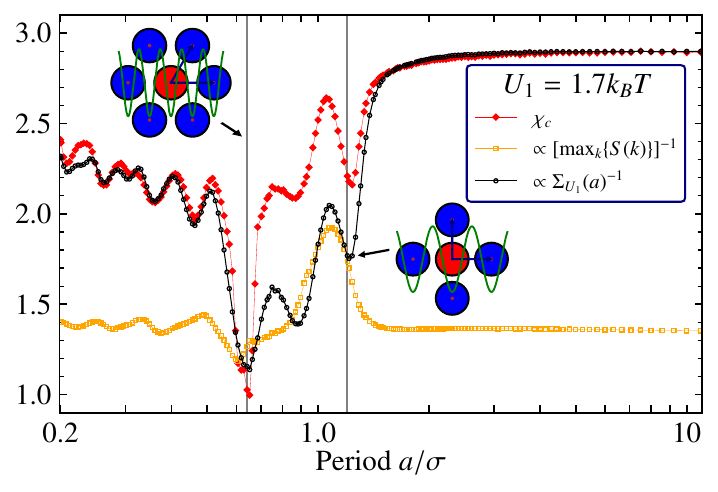}
\caption{\label{fig:var}  Comparison of  the phase-transition line $\chi_c= \chi_c(a, U_1; \bar{\varphi})$ to the behavior of the first peak of the angularly-averaged static structure factor $\max_k\{S(k)\}$ and the rescaled inverse variance of the static structure factor $\propto \Sigma_{U_1}(a)^{-1}$, for $\bar{\varphi} \approx 0.55$ and $U_1=1.7 k_B T$.
In correspondence of the minima of $\chi_c$, a sketch illustrates (on the left side) the preferred configurational hexagonal cage having a nearest-neighbor distance  fixed to match the packing fraction $\bar{\varphi}$. Two vertical gray lines are pointing to $a=0.64\sigma$ corresponding to a perfect hexagonal cage, and to $a=1.20\sigma$ for a perfect square cage.}
\end{figure}

The nonequilibrium-state diagram as calculated within MCT displays an oscillatory behavior of the glass-transition line with the modulation period $a$  in the range  $0.2 \lesssim a/\sigma \lesssim 2.0$, while bulk behavior is approached for $a/\sigma \gtrsim 2.0$, see 
Fig.~\ref{fig:phase}. Varying the modulation amplitude $0.5 \lesssim U_1/k_B T \lesssim 1.5 $, does not change the qualitative behavior of the glass-transition line, however, for larger $U_1/k_B T \gtrsim 1.5$ there is a preferred period $a/\sigma \approx 0.63$ emerging where the transition line is drastically changed by the external potential. For small periods,  the modulation generally promotes vitrification, while  for 
 $0.9\lesssim a/\sigma \lesssim 1.5$ and  $0.1 \lesssim U_1/k_B T \lesssim 1.0$ the external potential 
stabilizes the liquid phase  relative to the  glassy state, compare inset in Fig.~\ref{fig:phase}.

There is a striking correlation between the emergence of rims in the static structure factor and the oscillations of the glass-transition line (see also Supplemental Material~\cite{Ahmadirahmat_Supplemental:2025}).  
Within  MCT the state diagram is determined solely by the structure factors; conventionally the peaks in the structure factors are assumed to trigger the transition to structural arrest~\cite{Goetze:Complex_Dynamics}. 
Here, we show that not only the peaks of the static structure factor but also its minima play an important role.  
The maxima in the
static structure factors indicate the periodic distribution of particles,
while the minima reveal the regularity of the empty spaces between
them.
The caging effect, due to cage-like configurations of particles, emerges in supercooled liquids and becomes dominant in the glass~\cite{Li:Nature_587:2020}.  
Then, the variance of static structure factor should show how likely the cages form in supercooled liquids.
To quantify this insight, we define   the normalized variance of the static structure factor around its mean
\begin{align}
\Sigma_{U_1}(a) \equiv \frac{\Var[S_{U_1}(a)]}{\Var[S_{U_1=0}]} = \frac{\int \diff \vec{k} [S_{U_{1}}(\vec{k},a)-1]^2}{\int \diff \vec{k} [S_{U_{1}=0}(\vec{k})-1]^2} ,
\end{align}
and observe 
strong correlations with the glass transition line, see Fig.~\ref{fig:var}. More precisely, for periods below the preferred one, $a\lesssim 0.63\sigma$, rescaling the inverse variance
$\Sigma_{U_1}(a)^{-1}$  to match the bulk behavior ($a \gg \sigma $), serves as a quantitative proxy of the glass-transition line. We have checked that this observation holds also for other modulation amplitudes, $U_1$. 
  Therefore, we can intuitively rationalize the behavior of the glass-transition line in terms of the typical configurations of hard disks leading to the oscillations.
  
In our system, the hard-disk repulsion and external modulation are the driving forces leading to different particle configurations and eventually different static structure factors.
The interplay of these two forces promotes the formation of local cages of particles around other particles showing different shapes depending on the modulation period $a$.
In particular, for a period of the external modulation $a\approx 0.63 \sigma $, corresponding to the minimum of the glass transition line, we observe cages with a shape fluctuating around that of a hexagon such that one of its diameters is aligned with the modulation (see left-side sketch in Fig.~\ref{fig:var}). 
Indeed, the distance between particles in a hexagonal lattice having packing fraction $\bar{\varphi}=0.55$ is about $ 1.28 \sigma$ which is in a very good agreement with twice the period of the external modulation $a \approx 0.63 \sigma$. Similarly, locally square-shaped  cages form for the period of the external modulation $a\approx 1.22 \sigma $ (see right-side sketch in Fig.~\ref{fig:var}).
The most probable cages in 2D are hexagonal, square, and amorphous cages, and/or combination of all of them. Of course it is possible that for some periods the modulation will lower the caging effect.


\textit{Summary and conclusion.--} We have generalized MCT to encompass modulated liquids exposed to an external one-dimensional periodic potential.
The theory differs in several important aspects from the corresponding one in bulk (for details see companion paper \cite{Ahmadirahmat:PREL:2025}).

We have solved the MCT equations for the nonequilibrium state diagram of the system as a function of the period and amplitude of the modulation. Our results demonstrate several key findings regarding the behavior of colloidal liquids under external modulation. 
Firstly, we predict a multiple reentrant transition into the glassy state 
by merely 
varying the amplitude and period of the external potential, exploiting that specific combinations of these parameters can induce or suppress vitrification. 
Secondly, the system exhibits a preferred period where the glass transition becomes very sensitive to the amplitude of the modulation. For 
hard disks,  the critical parameter of the glass-transition line changes by almost a factor of three for moderate amplitudes in comparison to bulk.
Thirdly, for certain periods at low amplitude, the glassy state is suppressed (compared to a bulk colloidal liquid); a phenomenon analogous to the melting of colloidal crystals upon exposing them to external modulations~\cite{Chakrabarti:PRL_75:1995, Wei:PRL_81:1998,Strepp:PRE_63:2001, Bechinger:PRL_86:2001}.
However, there the crystal melting  occurs at high amplitudes, whereas in our case, it occurs also at lower  external potentials and is induced by cage distortion. 
All these behaviors are manifestations of  the promotion and suppression of the caging effect by the external potential.
The first observation also holds  for a colloidal suspension in confinement~\cite{Mandal:NatComm_5:2014}, while the other two are genuine  features of the modulated liquid.

Overall, these results predict that the phase of a colloidal liquid exposed to a periodic external potential can be effectively controlled by adjusting both the 
amplitude and period of the external potential, providing a robust framework for tuning the physical states of colloidal liquid with precision.

Our approach encompasses naturally also infinitesimal external perturbations and therefore by linear-response theory, higher-order correlation functions  become accessible similar to inhomogeneous mode-coupling theory~\cite{Biroli:PRL_97:2006}. 
Thus, it would be interesting to elaborate if 
the splitting of the currents employed here modifies some of the predictions for the higher-order susceptibilities.

\begin{acknowledgements}
We thank Rolf Schilling for constructive criticism on the manuscript. 
This research was funded in part by the Austrian Science Fund (FWF) 10.55776/I5257 and 10.55776/P35872.
\end{acknowledgements}

\end{document}